# DS-MTNet: Structured Multi-Task EEG Decoding for Human-Machine Collaboration

Xinjia Yu[a], Yang Zhou[a], Jing Yang[a], Tielin Shi[a], and Tao Cheng[b]*
[a]*School of Mechanical Science and Engineering, Huazhong University of Science and Technology, 1037 Luoyu Road, Wuhan, 430074, China*
[b]*School of Urban Transport and Logistics, Shenzhen Technology University, 3002 Lantian Road, Shenzhen, 518118, China*

***Abstract*** **—** Current human-machine collaboration (HMC) systems rely on environment-facing sensors to observe visible actions and scene states, but the internal perceptual, intention-related, and state-related processes of operators remain insufficiently integrated into machine perception. Electroencephalography (EEG) provides a non-invasive, time-resolved modality to capture neural activity associated with these processes and can serve as an additional sensing channel in HMC. However, HMC-relevant EEG evidence is often mixed in continuous recordings. Existing EEG decoding methods usually target task-specific classification or aggregate prediction, so multiple HMC-relevant readouts are rarely organized in a unified EEG representation. To address this gap, this paper proposed the Decomposed-Source Multi-Task Network (DS-MTNet), a structured multi-task EEG decoding framework. DS-MTNet integrated three streams, namely EEG waveforms, task-routed source embeddings, and temporal-spectral power features, into reusable slots and used dual gating mechanisms to route task-specific components. The model was tested on a sustained-attention driving EEG dataset with three representative readouts: lane-departure-related epochs for environmental-event processing, steering-response stage for response preparation, and reaction-time-defined alertness state for internal state. DS-MTNet achieved the best mean performance among traditional, single-task deep, and multi-task EEG baselines, with the most robust gains observed for steering-response stage decoding. Ablation and interpretability analyses suggested that DS-MTNet jointly decoded multiple readouts and organized event-related, response-related, and state-related EEG evidence in a unified source-slot representation. These findings provide a computational step toward incorporating operator-related neural evidence into machine perception in HMC.

***Index Terms*** **—** Electroencephalography (EEG), Human-machine Collaboration, Structured Representation Learning, Multi-task Learning, Cognitive Sensing

## I. INTRODUCTION

Human-machine collaboration (HMC) has expanded beyond industrial manufacturing tasks (e.g., human-robot collaborative assembly and disassembly) [1], [2] to open and ubiquitous settings such as autonomous driving [3], surgical assistance [4], and daily service robot applications [5], [6]. In these environments, safe and efficient collaboration requires robust bi-directional understanding between human operators and collaborative systems [7]. Through advanced user interfaces and physical feedback, humans can usually interpret the motion state and environmental context of machines. However, machine-side understanding of the human partner remains limited [8]. The current HMC systems do not sufficiently capture and use human-related information, particularly the perceptual and cognitive contributions of the operator to the collaborative loop.

This limitation mainly arises from the machine-side perception architecture used in the current HMC systems. These systems heavily rely on cameras, LiDAR, and other environment-facing sensors [8], [9]. Such sensors work well in structured environments but remain insufficient for complex human-machine interaction. They can capture only observable physical manifestations and cannot directly measure intrinsic human states or the physiological processes underlying decision-making [7]. Visible human motion is merely the final endpoint of complex neural processes such as perception, evaluation, and action preparation. Thus, the use of behavioral readouts misses the rich internal context of the operator. The current safety models are often built on rigid rules that treat humans as passive objects requiring protection [7], [10]. This perspective overlooks that human operators can act as active information sources whose neural activity may reflect multiple operator-related processes during collaboration.

Neural signals, particularly electroencephalography (EEG), provide a non-invasive and time-resolved window into neural activity associated with these internal processes [11]. During continuous interaction, a human operator may process perceptual information about the external environment [7], [12], form specific intentions [13], and experience fluctuations in endogenous states [14] . These concurrent cognitive and physiological processes are associated with corresponding neural responses. Because these varied processes arise in the same human perceptual-cognitive system, their neural correlates may be partially reflected in a common recording modality [15]. This shared access is one advantage of treating the human operator as an information source in HMC. EEG can capture these responses to a certain extent, but different forms of neural evidence are often mixed in the same continuous EEG signal [16]. Thus, decoding such mixed information may provide a useful computational basis for supplementing machine-side perception in complex HMC scenarios.

Interpreting this multi-dimensional neural information presents a specific computational challenge. EEG contains task-relevant evidence, but the corresponding features are highly multiplexed. Existing EEG decoding methods, including signal-processing pipelines and deep neural decoders, have achieved strong performance on many benchmark tasks. Much of this literature is still organized around task-specific classification [17], [18] or aggregate predictive metrics [19], and less attention has been paid to explicitly structuring multiple HMC-relevant readouts in a unified EEG representation framework.

To address this gap, this study developed the Decomposed-Source Multi-Task Network (DS-MTNet), a structured multi-task EEG decoding framework. We evaluated DS-MTNet on the sustained-attention driving task (SADT) dataset [20] and used three binary tasks as representative HMC-relevant readouts: lane-departure-related epochs (environmental-event-related information), steering-response stage (response-preparation information), and reaction-time-defined alertness state (internal-state information). DS-MTNet is designed to organize HMC-relevant EEG evidence into task-routed source-level representations. It integrates EEG waveforms, task-routed source embeddings, and temporal-spectral power features into reusable information slots, and uses gating mechanisms to select task-specific components. The experiments showed that DS-MTNet achieved the best mean performance among the traditional, single-task deep, and multi-task EEG baselines. The most statistically robust gains were observed for steering-response stage decoding. Together with ablation and interpretability analyses, these results suggest that DS-MTNet can jointly decode multiple HMC-relevant readouts and organize task-relevant EEG evidence within a unified source-slot representation. These findings provide an initial computational step toward EEG-assisted bi-directional sensing in HMC.

## II. BACKGROUND AND MOTIVATION

### *A. Environment-Facing Perception in HMC*

Perception modules in HMC are commonly used to estimate the states and relationships among humans, machines, task objects, and the external environment. Existing systems mainly rely on external physical and behavioral sensors, such as cameras, LiDAR, radar, inertial sensors, wearable sensors, vehicle sensors, robot sensors, and environmental sensing devices [9], [21]. Depending on the application, these sensors can provide information about scene structure, obstacles, machine state, human posture, gaze, action trajectories, human-machine distance, and risk events [22], [23], [24]. These signals support behavior recognition, action prediction, state monitoring, risk assessment, shared control, decision assistance, and adaptive collaboration [22], [23], [25].

This environment-facing perception paradigm has enabled many HMC applications, but its observations of the human operator are largely limited to overt behavioral manifestations or downstream action outcomes, leaving the underlying coupling among perception, intention, and endogenous state insufficiently modeled. For environmental events, this paradigm can capture physical changes but may not reliably determine whether the human operator has noticed or perceived them. Events that fall outside the predefined perception rules of the system may also remain undetected [7], [26]. These gaps may affect subsequent collaboration. For human intention, it mainly observes behavioral correlates instead of the formation of response preparation. Thus, environment-facings perception provides the foundation of HMC sensing, but it leaves room for complementary human-side sensing.

### *B. EEG-based Cognitive Sensing in HMC*

EEG is a suitable modality for cognitive sensing. Compared to behavioral, visual, or kinematic signals, EEG provides complementary evidence about internal processes that may not be directly observable from environment-facing physical sensors. Prior EEG studies have decoded neural patterns related to perception, attention, workload, fatigue, error awareness, and decision preparation [26], [27]. However, these EEG decoding efforts are often formulated as separate task-specific problems, so their outputs are rarely organized as multiple HMC-relevant readouts within one representation framework. This leaves open the problem of explicitly structuring multiple HMC-relevant readouts within a unified EEG representation framework.

This representation-level gap is further complicated by the intrinsic properties of EEG signals. EEG-based cognitive sensing remains challenging because EEG signals are noisy, non-stationary, subject-dependent, and artifact-prone [18]; EEG epochs may reflect different mixtures of event-related, response-related, and state-related factors. These properties motivate learning frameworks that combine complementary EEG representations and impose task-related structure, instead of relying on one flat feature representation.

### *C. Multi-view and Multi-task EEG Learning*

Multi-view EEG learning provides a technical basis for extracting complementary information from signals. EEG can be represented through temporal waveforms, spectral power, time-frequency patterns, spatial topology, functional connectivity, and source-domain features. These views reflect different neurophysiological properties and have motivated EEG learning methods based on feature fusion, attention-based integration, graph representation, and contrastive alignment [28], [29]. For cognitive sensing in HMC, this multi-view property is important because environmental-event-related processing, response preparation, and internal-state information may not be expressed in one feature view alone. Many multi-view approaches mainly concatenate or fuse features for classification, and they do not explicitly organize the fused evidence into reusable information components that can be read by different tasks.

Multi-task learning provides another basis for cognitive sensing because several HMC-relevant targets may be related. Instead of training an independent model for each cognitive label, multi-task models usually share part of the representation space and use task-specific branches, heads, gates, or expert modules to preserve task-dependent patterns. This design is useful for EEG because labeled samples are often limited, subject variability is high, and different cognitive responses may share neural substrates. However, conventional multi-task EEG models may still suffer from negative transfer, task imbalance, and insufficient separation between shared and task-specific components. Many models use an undifferentiated shared backbone and split only near the output layer, which makes it difficult to determine which information is shared, which information is task-specific, and which EEG view or frequency band supports each task [30], [31], [32]. These limitations motivate a structured multi-task EEG framework for HMC-oriented cognitive sensing. Such a framework should not only improve classification performance, but also organize EEG representations into reusable components and allow each task to read out task-relevant information. The next section presents DS-MTNet as one implementation of this design principle.

## III. PROPOSED DS-MTNET MODEL

### *A. Overview and Problem Formulation*

DS-MTNet is designed to decode multiple operator-related labels from one EEG window in HMC; the signal may contain mixed evidence related to environmental events, response preparation, and internal state. The EEG window is $\mathbf{X} \in \mathbb{R}^{C\times T}$, where $C$ is the number of EEG channels and $T$ is the number of time samples. Unless otherwise stated, the batch dimension is omitted for clarity.

**Fig. 1** summarizes the forward process of DS-MTNet. The model receives one preprocessed EEG window $\mathbf{X}$ and outputs one logit vector $\mathbf{o}_{\text{final}}^{(m)} \in \mathbb{R}^{n_m}$ for each task $m$. The source branch first converts $\mathbf{X}$ into a shared dictionary of frequency-constrained candidate sources, routes this dictionary into task-specific source slots, and produces source logits $\mathbf{o}_{\text{src}}^{(m)}$. In parallel, the information branch receives three views derived from the same window: the normalized EEG, the task-source embeddings, and the candidate-source power representation. It decomposes these views into reusable information slots and produces information logits $\mathbf{o}_{\text{info}}^{(m)}$. The final task head fuses the two logit vectors using a non-negative task gate, so the source branch remains the direct decoder and the information branch supplies complementary multi-view evidence.

Let $M$ be the number of tasks and $n_m$ be the number of classes for task $m$. DS-MTNet maps each valid EEG window to the labels of the corresponding tasks. The supervised loss is computed only for tasks with valid annotations; when task $m$ is valid, its label is $y^{(m)} \in \{1, \dots, n_m\}$. Section IV reports the values of $C$, $T$, $M$, and $n_m$ in the SADT experiments.

### *B. Source Branch*

The source branch receives the preprocessed EEG window $\mathbf{X} \in \mathbb{R}^{C\times T}$ and outputs task-specific source logits $\{\mathbf{o}_{\text{src}}^{(m)}\}_{m=1}^{M}$. Its intermediate output is a shared source dictionary $\mathbf{S} \in \mathbb{R}^{F\times K\times T}$, where $F$ is the number of frequency bands and $K$ is the number of candidate sources per band. This design separates the learning of a common source space from the task-specific selection of source evidence.

Before the learnable source decomposition, the EEG window is normalized channel by channel. For each channel $c$ and time sample $t$, the training-fold mean $\mu_c$ and standard deviation $\sigma_c$ define the normalized window $\widetilde{\mathbf{X}} \in \mathbb{R}^{C\times T}$:

$$\tilde{X}_{c,t} = \frac{X_{c,t} - \mu_c}{\sigma_c + \varepsilon}, (1)$$

where $\varepsilon$ is a small constant for numerical stability. This step keeps the tensor shape unchanged, $\mathbf{X} \to \widetilde{\mathbf{X}}$, and places all channels on a comparable numerical scale. The statistics are estimated only from the training fold and then applied to the validation and test splits, to avoid leakage from held-out data.

The normalized window is first transformed by a learnable spatial whitening layer, $\mathbf{X}_w = \mathbf{W}_w\widetilde{\mathbf{X}}$, where $\mathbf{W}_w \in \mathbb{R}^{C\times C}$. The output $\mathbf{X}_w \in \mathbb{R}^{C\times T}$ preserves the original time axis but conditions the channel space for source separation. The whitening matrix is initialized from the training-fold covariance and is then optimized with the rest of the network. This layer follows the intuition of independent component analysis (ICA)-style [33] preprocessing: before learning candidate source components, the observed sensor channels should be placed in a better-conditioned coordinate system.

A parallel SincNet [34] filter bank then receives $\mathbf{X}_w$ and outputs $F$ band-limited tensors $\{\mathbf{X}_b\}_{b=1}^{F}$, each in $\mathbb{R}^{C\times T}$. Each Sinc filter is shared across channels, so the module acts as a frequency-selective temporal filter. The bands are initialized to 1-4, 4-8, 8-13, 13-30, and 30-100 Hz. These initial bands follow common EEG frequency ranges, but the cutoff frequencies remain learnable under lower-bound constraints. Therefore, the filter bank gives the model a physiologically meaningful starting point while still allowing data-driven adjustment.

The band-wise unmixing layer receives each band-limited tensor $\mathbf{X}_b$ and outputs $K$ candidate source signals for that band. The operation and resulting dictionary are

$$\mathbf{S}_b(:,t) = \mathbf{W}_b\mathbf{X}_b(:,t), \mathbf{S} = \{S_{b,k}\} \in \mathbb{R}^{F\times K\times T}. (2)$$

Here, $\mathbf{W}_b \in \mathbb{R}^{K\times C}$ is the learnable unmixing matrix for band $b$. The output $\mathbf{S}_b \in \mathbb{R}^{K\times T}$ contains $K$ candidate temporal sources in that frequency band, and stacking all bands forms the dictionary $\mathbf{S}$. This dictionary is shared by all tasks, serves as the input to the task router, and also provides the power-view input for the information branch.

The task router receives the complete dictionary $\mathbf{S}$ and outputs task-routed sources $\widetilde{\mathbf{S}} \in \mathbb{R}^{M\times R\times T}$. For each task $m$ and source slot $r$, the router learns a softmax distribution $\alpha_{m,r}$ over all $F \times K$ candidate sources. The routed source is

$$\tilde{S}_{m,r}(t)=\sum_{b=1}^{F}\sum_{k=1}^{K}\alpha_{m,r,b,k}\,S_{b,k}(t),\alpha_{m,r}=\mathrm{softmax}(a_{m,r}).\ (3)$$

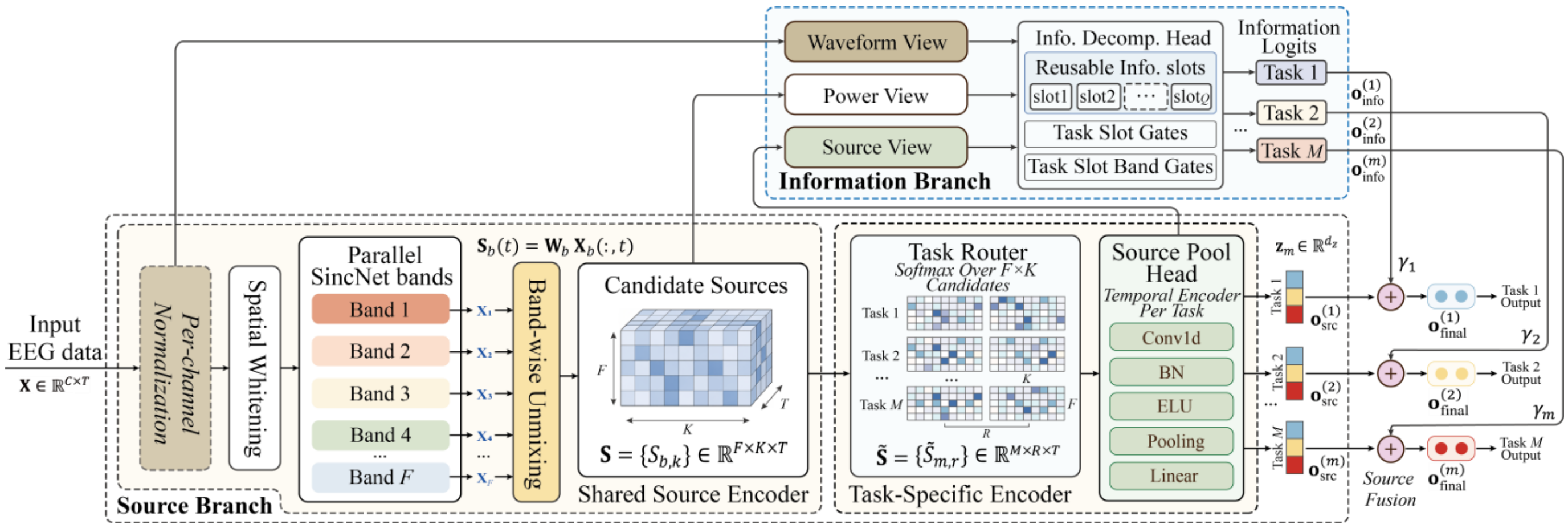

**Fig. 1.** Overall architecture of DS-MTNet.

Thus, each source slot is a soft mixture over the full candidate dictionary instead of a hard selection of one source. This is important because task-relevant EEG evidence may not correspond to one single isolated candidate source. The router therefore reduces the shared $F\times K$ dictionary to a compact task-specific representation with shape $\mathbb{R}^{R\times T}$ for each task.

The SourcePoolHead receives the routed sources of one task, $\tilde{\mathbf{S}}_m\in\mathbb{R}^{R\times T}$, and outputs two quantities: a source embedding $\mathbf{z}_m\in\mathbb{R}^{d_z}$ and source logits $\mathbf{o}_{\mathrm{src}}^{(m)}\in\mathbb{R}^{n_m}$. The temporal encoder uses two 1-D convolution blocks. The first block maps the $R$ routed source slots to 16 temporal feature maps with kernel size 15 and is followed by batch normalization, ELU activation, and average pooling with size 4. The second block keeps 16 temporal feature maps with the same kernel size and applies batch normalization and ELU activation. The embedding $\mathbf{z}_m=f_{\mathrm{src},m}(\tilde{\mathbf{S}}_m)$ is used by the information branch as the source view, and a task-specific linear classifier maps $\mathbf{z}_m$ to $\mathbf{o}_{\mathrm{src}}^{(m)}$.

### *C. Information Branch*

The information branch receives three inputs derived from the same EEG window and outputs one information logit vector $\mathbf{o}_{\mathrm{info}}^{(m)}\in\mathbb{R}^{n_m}$ for each task. As shown in **Fig. 2**, its inputs are the normalized EEG $\tilde{\mathbf{X}}\in\mathbb{R}^{C\times T}$, the task source embeddings $\{\mathbf{z}_m\}_{m=1}^{M}$, and the candidate-source dictionary $\mathbf{S}\in\mathbb{R}^{F\times K\times T}$. These inputs form waveform, source, and power views, respectively. The purpose of this branch is to collect complementary evidence that may be distributed across the normalized EEG waveform, task-routed source embeddings, and frequency-power patterns.

The three view encoders map heterogeneous inputs to the same information dimension $d_h$. The waveform-view encoder $\psi_{\mathrm{wav}}$ is an EEGNet-style [35] temporal-spatial backbone applied to $\tilde{\mathbf{X}}$. It provides a compact representation and retains information that may be weakened by source routing. The source-view projector $\psi_{\mathrm{src}}$ receives the concatenated source embeddings $\mathbf{z}_{\mathrm{src}}=[\mathbf{z}_1;\ldots;\mathbf{z}_M]\in\mathbb{R}^{M\cdot d_z}$, so the information branch can reuse the task-routed evidence learned by the source branch. The power-view encoder receives time-binned log-variance features computed from $\mathbf{S}$ and outputs a global power feature and per-band power features. The resulting global view features are $\mathbf{v}_{\mathrm{wav}},\mathbf{v}_{\mathrm{src}},\mathbf{v}_{\mathrm{pow}}\in\mathbb{R}^{d_h}$, and the per-band features are $\{\mathbf{p}_b\}_{b=1}^{F}$, with $\mathbf{p}_b\in\mathbb{R}^{d_h}$.

The global view features are stacked as $\mathbf{V}=[\mathbf{v}_{\mathrm{wav}};\mathbf{v}_{\mathrm{src}};\mathbf{v}_{\mathrm{pow}}]\in\mathbb{R}^{3\times d_h}$. The slot-view gate receives $\mathbf{V}$ and produces $Q$ reusable information slots. For slot $q$, the gate $\boldsymbol{\pi}_q=\mathrm{softmax}(\mathbf{u}_q)\in\mathbb{R}^3$ weights the waveform, source, and power views, and a slot-specific transform produces the shared slot:

$$\mathbf{Z}_q=\phi_q(\boldsymbol{\pi}_q^{\top}\mathbf{V})\in\mathbb{R}^{d_h}.\quad(4)$$

The slots are shared across tasks; the model learns their roles from data. A slot may therefore represent a different weighted combination of temporal-spatial waveform, source-level task evidence, and power evidence.

The task-specific readout uses two gates. The task-slot-band gate $\boldsymbol{\eta}_{m,q}\in\mathbb{R}^F$ selects band-specific power evidence for task $m$ and slot $q$, and the task-slot gate $\boldsymbol{\rho}_m\in\mathbb{R}^Q$ selects among the adapted slots. The adapted slot and the final information feature are

$$\tilde{\mathbf{Z}}_{m,q}=\mathbf{Z}_q+\lambda_{\mathrm{band}}\sum_{b=1}^{F}\eta_{m,q,b}\,\mathbf{p}_b,\mathbf{h}_{\mathrm{info}}^{(m)}=\sum_{q=1}^{Q}\rho_{m,q}\,\tilde{\mathbf{Z}}_{m,q}.\quad(5)$$

The task-slot-band gate lets the same reusable slot respond differently to different frequency bands. This is useful because task-relevant information may be expressed through different spectral components for different readout tasks. The task-slot gate then summarizes the adapted slots into one task-specific information feature $\mathbf{h}_{\mathrm{info}}^{(m)}\in\mathbb{R}^{d_h}$. A task-specific linear head maps this feature to $\mathbf{o}_{\mathrm{info}}^{(m)}\in\mathbb{R}^{n_m}$. These logits do not serve as stand-alone predictions; the final fusion head combines them with the source logits.

### *D. Logit Fusion and Training Objective*

The fusion head receives two logit vectors for each task: the source logits $\mathbf{o}_{\mathrm{src}}^{(m)}\in\mathbb{R}^{n_m}$ and the information logits $\mathbf{o}_{\mathrm{info}}^{(m)}\in\mathbb{R}^{n_m}$. It outputs the final task logits through a non-negative gate:

$$\mathbf{o}_{\mathrm{final}}^{(m)}=\mathbf{o}_{\mathrm{src}}^{(m)}+\mathrm{softplus}(r_m)\mathbf{o}_{\mathrm{info}}^{(m)}.\ (6)$$

Here, $r_m$ is a learnable scalar for task $m$, and the softplus function constrains the fusion weight to be non-negative. The posterior probability for task $m$ is $\mathbf{p}^{(m)}=\mathrm{softmax}(\mathbf{o}_{\mathrm{final}}^{(m)})$. This fusion keeps the source branch as the direct decoder and lets the information branch contribute only through a learned task-specific gate. Compared with feature concatenation, logit-level fusion also keeps the two branches' contributions separable for inspection: $\mathbf{o}_{\text{src}}^{(m)}$ reflects source-based evidence, whereas $\mathbf{o}_{\text{info}}^{(m)}$ reflects evidence from reusable multi-view information slots.

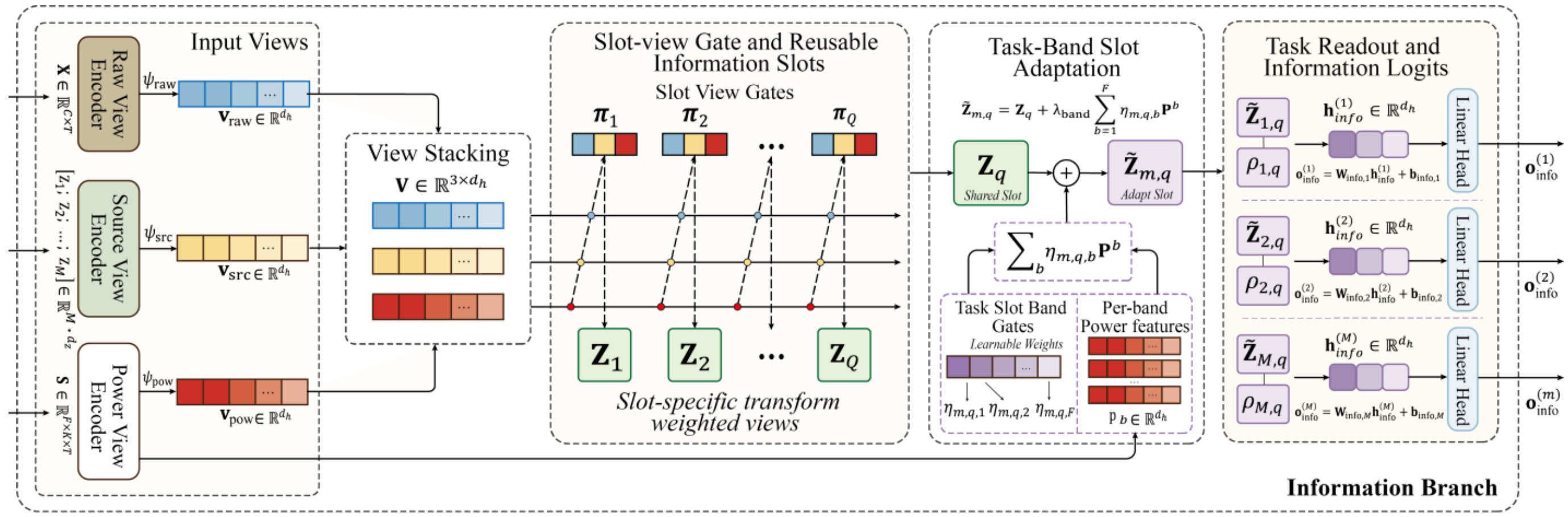

**Fig. 2.** Detailed information branch of DS-MTNet.

The supervised term $\mathcal{L}_{\text{task}}$ aggregates the cross-entropy losses of all task heads using their final logits, with each task evaluated only on samples that have valid annotations. The full objective is

$$\mathcal{L} = \mathcal{L}_{\text{task}} + \lambda_{\text{rec}}\mathcal{L}_{\text{rec}} + \lambda_{\text{corr}}\mathcal{L}_{\text{corr}} + \lambda_{\text{sparse}}\mathcal{L}_{\text{sparse}}. \quad (7)$$

The three regularizers act on the source decomposition. The reconstruction loss $\mathcal{L}_{\text{rec}}$ linearly reconstructs the whitened EEG $\mathbf{X}_w$ from the complete source dictionary $\mathbf{S}$, encouraging the candidate sources to retain signal information beyond the supervised labels. The correlation loss $\mathcal{L}_{\text{corr}}$ penalizes high correlation between routed task sources, encouraging the selected source slots to carry less redundant evidence. The sparsity loss $\mathcal{L}_{\text{sparse}}$ encourages sharper routing distributions and makes the task-source associations easier to interpret. During inference, the reconstruction output is not used for prediction, and no reconstruction or regularization losses are evaluated.

## IV. EXPERIMENT SETUP

### *A. Dataset, Tasks, and Evaluation Protocol*

We evaluated DS-MTNet on the public SADT EEG dataset reported by Cao et al. [20]. The original database contained 62 sessions from 27 subjects in an immersive virtual-reality driving simulator. To keep the kinesthetic condition consistent, we used only the motion-enabled sessions, yielding 28 sessions from 20 subjects.

The recordings were sampled at 500 Hz and included 30 EEG channels, two mastoid reference electrodes, and one vehicle-position channel. The EEG was re-referenced to linked mastoids, band-pass filtered at 0.5–100 Hz, notch-filtered at 60 Hz, and segmented into 3-s windows with T=1500 samples. After artifact screening, the final cache contained 45,006 usable EEG epochs.

We defined M=3 binary tasks from the event markers and behavioral reaction time. Task 1 was lane-departure detection: positive epochs were extracted from [-1, +2] s around the deviation onset, whereas negative epochs were extracted from the post-response rest period [+0.5, +3.5] s after response offset and were discarded if they overlapped with the next deviation event. Task 1 was therefore treated as a representative task for environmental-event-related EEG information. Task 2 was steering-response stage decoding: preparation epochs were extracted from [-3, 0] s before response onset, and action-onset epochs were extracted from [-1, +2] s around response onset. This task was used to test whether short-window EEG could distinguish response-stage-related activity associated with steering preparation and action onset. Task 3 was reaction-time-defined alertness-state decoding: epochs were extracted from the pre-deviation interval [-3, 0] s before deviation onset, and labels were defined in each session according to reaction time. The 25% of trials with the shortest reaction times were treated as high-alertness states, whereas the 25% of trials with the longest reaction times were treated as low-alertness states. The valid label counts were 20,332 for Task 1, 19,905 for Task 2, and 4,769 for Task 3.

We used a 20-fold leave-one-subject-out protocol to evaluate cross-subject generalization. In each fold, one subject was held out for testing, and the remaining subjects were split into training and internal validation sets. When a held-out subject had multiple sessions, all sessions from that subject were assigned to the test fold. Channel-wise normalization statistics were estimated only from the training subjects and then applied to the validation and test data. The proposed and baseline models used the same epochs, preprocessing pipeline, label definitions, and subject splits.

### *B. Implementation, Model Settings, and Baselines*

All deep-learning models were implemented in PyTorch and trained on NVIDIA RTX 4090 GPUs. DS-MTNet used $C = 30$, $T = 1500$, $M = 3$, and $n_1 = n_2 = n_3 = 2$. The source branch used $F = 5$ SincNet bands initialized to 1–4, 4–8, 8–13, 13–30, and 30–100 Hz, with kernel lengths of 251, 125, 63, 31, and 31 samples. Each band produced $K = 8$ candidate sources, and the task router produced $R = 2$ routed source slots per task. The source embedding dimension was $d_z = 64$. The information branch used $Q = 4$ reusable slots, an information dimension $d_h = 32$, and $U = 6$ temporal bins for the power view.

The models were trained with Adam using a learning rate of 0.001, a weight decay of $5.0 \times 10^{-5}$, and a batch size of 64. The maximum number of epochs was 150, and early stopping used a patience of 30 epochs based on internal validation performance. The supervised loss was computed with task-wise cross-entropy over valid labels. The source regularization weights were $\lambda_{\text{rec}} = 0.005$, $\lambda_{\text{corr}} = 0.005$, and $\lambda_{\text{sparse}} = 0.001$.

We compared DS-MTNet with traditional, single-task deep, and multi-task deep baselines. The traditional baselines were CSP-SVM [36],[37] and FBCSP-SVM [38]. The single-task deep baselines were EEGNet [35], FBCNet [39], ShallowConvNet [40], EEG Conformer [29], and ATCNet [18]; each was trained separately for the three tasks and reported as × 3. The multi-task baselines were multi-task EEGNet, MMoE+EEGNet [18], [32], and CGC+EEGNet [31]. These baselines tested whether conventional spatial filtering, single-task temporal-spatial decoding, or standard multi-task sharing could match the proposed frequency-constrained source decomposition and information-slot fusion.

## V. RESULTS

### *A. Performance*

Table I shows the comparison between DS-MTNet and the baseline methods on the SADT dataset. Using a single jointly trained model, DS-MTNet attained the highest mean ACC and F1 in all three tasks. It obtained 81.70% ACC and 81.31% F1 for Task 1, 85.12% ACC and 85.09% F1 for Task 2, and 77.60% ACC and 76.31% F1 for Task 3. Compared to the best baseline in each task, DS-MTNet improved the ACC/F1 scores by 1.53/1.48 percentage points for Task 1, 2.65/2.73 percentage points for Task 2, and 1.14/1.64 percentage points for Task 3. These results provide performance-level evidence that one structured model can jointly decode three related but heterogeneous EEG readouts, the strength of the advantage differed across tasks.

The traditional methods showed lower performance, especially in Task 1 and Task 2. FBCSP-SVM performed better than CSP-SVM, which is consistent with the use of band-wise spatial features. However, it was still lower than DS-MTNet by 19.22/22.25 percentage points in ACC/F1 for Task 1 and by 23.50/26.95 percentage points in ACC/F1 for Task 2. This result suggests that hand-crafted spatial-variance features alone may not fully represent the task-related EEG dynamics in the SADT dataset.

The single-task deep-learning baselines outperformed the traditional methods. FBCNet achieved the best single-task result in Task 1, whereas ATCNet achieved the best single-task result in Task 2. DS-MTNet still obtained higher ACC and F1 values than these models. The improvement over FBCNet was 1.77/1.69 percentage points in ACC/F1 for Task 1, and the improvement over ATCNet was 3.50/3.80 percentage points in ACC/F1 for Task 2. This result suggests that training each task independently may not fully use the shared information among related cognitive tasks.

Among the multi-task baselines, CGC+EEGNet was the strongest, achieving 80.17%/79.83%, 82.47%/82.36%, and 76.46%/74.67% ACC/F1 for Tasks 1-3, respectively. DS-MTNet achieved higher ACC and F1 on all three tasks, showing that it was the strongest overall multi-task model under the same SADT evaluation setting.

### *B. Statistical Robustness Across LOSO Folds*

To assess cross-subject robustness, we performed paired LOSO-fold comparisons between DS-MTNet and the representative baselines, using each subject as the statistical unit. Table II summarizes the FDR-corrected paired comparisons for ACC. With significance set at $q < 0.05$, DS-MTNet showed significant gains over EEGNet ×3, ATCNet ×3, and Multi-task EEGNet on Tasks 1 and 2 (all $q < 0.02$). Against Multi-task EEGNet, the improvement was positive in 18/20 folds for Task 1 and 17/20 folds for Task 2, which indicates that these gains were not driven by a small number of subjects.

Against the stronger expert-sharing multi-task baselines, the advantage was more selective. DS-MTNet remained significantly better than MMoE+EEGNet on Tasks 2 and 3 and better than CGC+EEGNet on Task 2. The differences against CGC+EEGNet were positive but not significant for Tasks 1 and 3. Thus, the most statistically robust improvement appeared in Task 2, where DS-MTNet achieved significant gains against all five representative baselines. This result suggests that the overall DS-MTNet design provided the most consistent benefit for steering-response stage decoding, whereas the Task 3 advantage was less consistent against the strongest expert-sharing baseline.

TABLE I
COMPARISON OF DS-MTNET WITH BASELINE METHODS ON THE SADT DATASET (MEAN ± SD, %)

| **Methods** | **Task 1** | | **Task 2** | | **Task 3** | |
|---|---|---|---|---|---|---|
| | ACC | F1 | ACC | F1 | ACC | F1 |
| Traditional methods | | | | | | |
| CSP-SVM | 57.35 ± 5.14 | 53.97 ± 8.38 | 57.65 ± 5.19 | 52.28 ± 10.33 | 74.86 ± 9.91 | 73.26 ± 13.04 |
| FBCSP-SVM | 62.48 ± 6.59 | 59.06 ± 9.36 | 61.62 ± 6.70 | 58.14 ± 10.59 | 75.17 ± 9.27 | 73.55 ± 12.48 |
| Single-task methods | | | | | | |
| EEGNet × 3 | 77.56 ± 7.59 | 77.20 ± 7.85 | 81.38 ± 7.41 | 81.25 ± 7.52 | 75.73 ± 10.17 | 74.03 ± 13.39 |
| ShallowConvNet×3 | 76.80 ± 8.07 | 76.18 ± 8.84 | 80.14 ± 7.53 | 79.98 ± 7.67 | 74.34 ± 9.84 | 72.72 ± 12.72 |
| FBCNet ×3 | 79.93 ± 6.73 | 79.62 ± 7.00 | 79.67 ± 7.05 | 79.17 ± 7.60 | 75.34 ± 9.41 | 73.67 ± 12.46 |
| EEG Conformer ×3 | 69.63 ± 8.06 | 68.37 ± 9.46 | 67.62 ± 7.90 | 65.46 ± 10.19 | 75.89 ± 10.34 | 74.13 ± 13.58 |
| ATCNet ×3 | 76.97 ± 9.06 | 76.11 ± 10.14 | 81.62 ± 7.43 | 81.29 ± 7.96 | 75.26 ± 10.07 | 73.58 ± 13.27 |
| Multi-task methods | | | | | | |
| Multi-task EEGNet | 77.35 ± 6.14 | 77.12 ± 6.31 | 80.81 ± 7.10 | 80.70 ± 7.19 | 74.78 ± 9.42 | 73.00 ± 12.84 |
| MMoE+EEGNet | 79.29 ± 8.45 | 78.93 ± 8.94 | 82.24 ± 6.70 | 82.08 ± 6.94 | 73.53 ± 10.07 | 71.44 ± 13.47 |
| CGC+EEGNet | 80.17 ± 8.42 | 79.83 ± 8.85 | 82.47 ± 6.60 | 82.36 ± 6.72 | 76.46 ± 11.13 | 74.67 ± 14.21 |
| **DS-MTNet** | **81.70 ± 7.09** | **81.31 ± 7.67** | **85.12 ± 5.72** | **85.09 ± 5.73** | **77.60 ± 11.43** | **76.31 ± 13.53** |

TABLE II
PAIRED LOSO ROBUSTNESS OF DS-MTNET AGAINST REPRESENTATIVE BASELINES

| Baseline | Task 1 ΔACC / q / n+/n0/n− | Task 2 ΔACC / q / n+/n0/n− | Task 3 ΔACC / q / n+/n0/n− |
|---|---|---|---|
| EEGNet ×3 | +4.14 / 0.0004 / 18/0/2 | +3.73 / 0.0024 / 15/0/5 | +1.87 / 0.1812 / 13/1/6 |
| ATCNet ×3 | +4.73 / 0.0052 / 15/0/5 | +3.50 / 0.0161 / 15/0/5 | +2.34 / 0.1903 / 14/0/6 |
| Multi-task EEGNet | +4.35 / 0.0011 / 18/0/2 | +4.31 / 0.0011 / 17/0/3 | +2.83 / 0.0592 / 14/2/4 |
| MMoE+EEGNet | +2.40 / 0.1179 / 14/0/6 | +2.88 / 0.0430 / 14/1/5 | +4.08 / 0.0200 / 15/0/5 |
| CGC+EEGNet | +1.53 / 0.3153 / 12/0/8 | +2.65 / 0.0270 / 15/0/5 | +1.14 / 0.3927 / 9/2/9 |

### *C. Ablation Studies*

The ablation study tested the hypothesis that the performance of DS-MTNet stemmed from its structured decomposition, in which gates specific to each task read out a shared and reusable representation, instead of from generic capacity for multiple tasks. We removed or simplified each structural component in turn and re-evaluated under the same 20-fold subject-level LOSO protocol. Table III reports the ACC/F1 of the full model and eight ablated variants. The full DS-MTNet obtained the best overall result (mean ACC/F1 = 81.47%/80.90%), and every ablation reduced mean performance, suggesting that each component was useful in the full configuration.

*Shared structured representation.* Removing the reusable information slots produced the largest degradation by a wide margin (mean ACC/F1 = 77.30%/76.62%, a reduction of 4.17/4.28 points). In this variant the three views were concatenated and classified directly, so the drop mainly reflected the value of organizing multi-view EEG into reusable slots before read-out. This was the strongest ablation-level support for the structured representation path. The loss concentrated on Task 2, with a 6.26-point ACC drop, which suggests that the slot structure played an important role in the response-stage-related readout. This pattern was consistent with the LOSO-fold analysis, where Task 2 showed the most statistically robust improvement across representative baselines. Disabling the source decomposition (SincNet band-pass filtering, band-wise unmixing and candidate source dictionary construction), in which the sources were held fixed instead of learned, lowered the mean ACC/F1 to 79.59%/79.10% (a reduction of 1.88/1.80 points). This suggests that the learned source representation supplied informative intermediate features for the downstream views and read-out.

*Task-specific read-out.* Replacing the task-specific source routing (task router) with one shared source readout reduced the mean ACC/F1 to 80.12%/79.49% (a reduction of 1.35/1.41 points), so one common source mixture for all tasks was less effective than letting each task select task-relevant sources. The two slot-level gates showed the same pattern but with a sharper interpretation. Replacing the task-slot gate with equal slot weights cost 1.62/1.62 points, and removing the frequency (band) gate cost 1.28/1.35 points on average. The frequency gate, however, was almost entirely specific to Task 3. It left Task 1 and Task 2 nearly unchanged (a drop of 0.29/0.84 ACC) but accounted for a 2.71-point ACC drop on Task 3. This selectivity was consistent with the common observation that alertness-related EEG activities are expressed through frequency-specific spectral changes [41].

*Multi-view input.* The waveform, source, and power views each contributed complementary information. Removing the source view caused the largest view-level drop (mean ACC/F1 = 79.14%/78.48%, a reduction of 2.33/2.42 points), followed by the power view (79.49%/78.90%, 1.98/2.00 points) and the waveform view (80.15%/79.49%, 1.32/1.41 points). The source view therefore had the largest effect among the three views in this ablation. Two regularities emerged from the ablation results. First, the largest losses occurred after removing the reusable information slots and the source-related pathway, whereas the task-specific gates and the waveform view produced smaller reductions. This pattern supports the interpretation that the performance of DS-MTNet was related to the structured shared-plus-specific decomposition, not only to the presence of multiple task heads. Second, Task 3 was the most sensitive target across most ablations and was the only task with a large drop after removing the frequency gate. This suggests that the reaction-time-defined alertness readout depended more on the full decomposition and frequency-resolved gating than the other two tasks. Overall, the source decomposition, tri-view representation, reusable information slots, and gated task-specific read-out appeared to act together, and their combination formed the most effective DS-MTNet configuration.

### *D. Information Decomposition Analysis*

**Fig. 3** shows how DS-MTNet organized the three decoding tasks at two levels: task-specific source routing and task-specific information-slot reading. Both maps were averaged across the 20 LOSO folds.

**Fig. 3 (a)** shows the task-specific use of frequency bands and learned band-wise candidate sources. For each task, the router weights were normalized over the same 5 × 8 set of candidate sources, so the equal-use reference was 2.50%. The three tasks showed weak but reproducible band-level preferences. Fold-wise tests showed that Task 1 assigned significantly higher weight to beta-band candidates, Task 2 to theta-band candidates, and Task 3 to gamma- and theta-band candidates after FDR correction. Descriptively, Task 3 showed a more distributed pattern than Tasks 1 and 2, with weights spread across several bands. The between-task differences were small, with most weights between about 2.0% and 3.1%, so this map gives only descriptive evidence of task-dependent source-frequency routing.

TABLE III
ABLATION STUDY OF DS-MTNET ON THE SADT DATASET

| Model Variant | Ablated | Task 1 | Task 2 | Task 3 | Mean Acc/F1 | ΔAcc/F1 |
|---|---|---|---|---|---|---|
| **Full** | **None** | **81.70/81.31** | **85.12/85.09** | **77.60/76.31** | **81.47/80.90** | **0.00/0.00** |
| SrcDec | No source decomposition | 78.28/77.84 | 83.14/83.07 | 77.36/76.40 | 79.59/79.10 | 1.88/1.80 |
| SrcRoute | Shared source readout | 80.82/80.31 | 84.09/84.05 | 75.44/74.12 | 80.12/79.49 | 1.35/1.41 |
| WavView | No waveform view | 81.17/80.87 | 83.88/83.82 | 75.39/73.78 | 80.15/79.49 | 1.32/1.41 |
| SrcView | No source view | 79.23/78.49 | 82.95/82.82 | 75.25/74.12 | 79.14/78.48 | 2.33/2.42 |
| PowView | No power view | 79.37/79.10 | 82.89/82.85 | 76.20/74.74 | 79.49/78.90 | 1.98/2.00 |
| InfoSlot | No information slots | 77.58/77.29 | 78.86/78.68 | 75.46/73.89 | 77.30/76.62 | 4.17/4.28 |
| SlotGate | Equal slot weights | 80.31/79.98 | 83.52/83.46 | 75.73/74.39 | 79.85/79.28 | 1.62/1.62 |
| FreqGate | No frequency gate | 81.41/81.19 | 84.28/84.24 | 74.89/73.22 | 80.19/79.55 | 1.28/1.35 |

**Fig. 3(b)** shows a clearer task-specific allocation at the information-slot level. Slot 1 received Task 1 (70.63 ± 1.94%), slot 2 dominated Task 2 (72.61 ± 1.85%), and slot 3 dominated Task 3 (70.55 ± 1.36%). The off-diagonal weights were much lower, between 7.81% and 11.96% on average, and the residual slot received a small and similar weight from every task (10.4% to 12.0%). Each diagonal weight was above the uniform 25% reference (one-sample t-tests across the 20 folds, FDR-corrected; all $p < 0.05$). The task-slot gate was initialized so that each task started from one slot, but the initialization alone was unlikely to account for the result, because the equal-slot-weight ablation (Table III) reduced the mean ACC/F1 by 1.62/1.62 points. Thus, the learned task-slot allocation appeared to have a functional role instead of being only a visualization artifact.

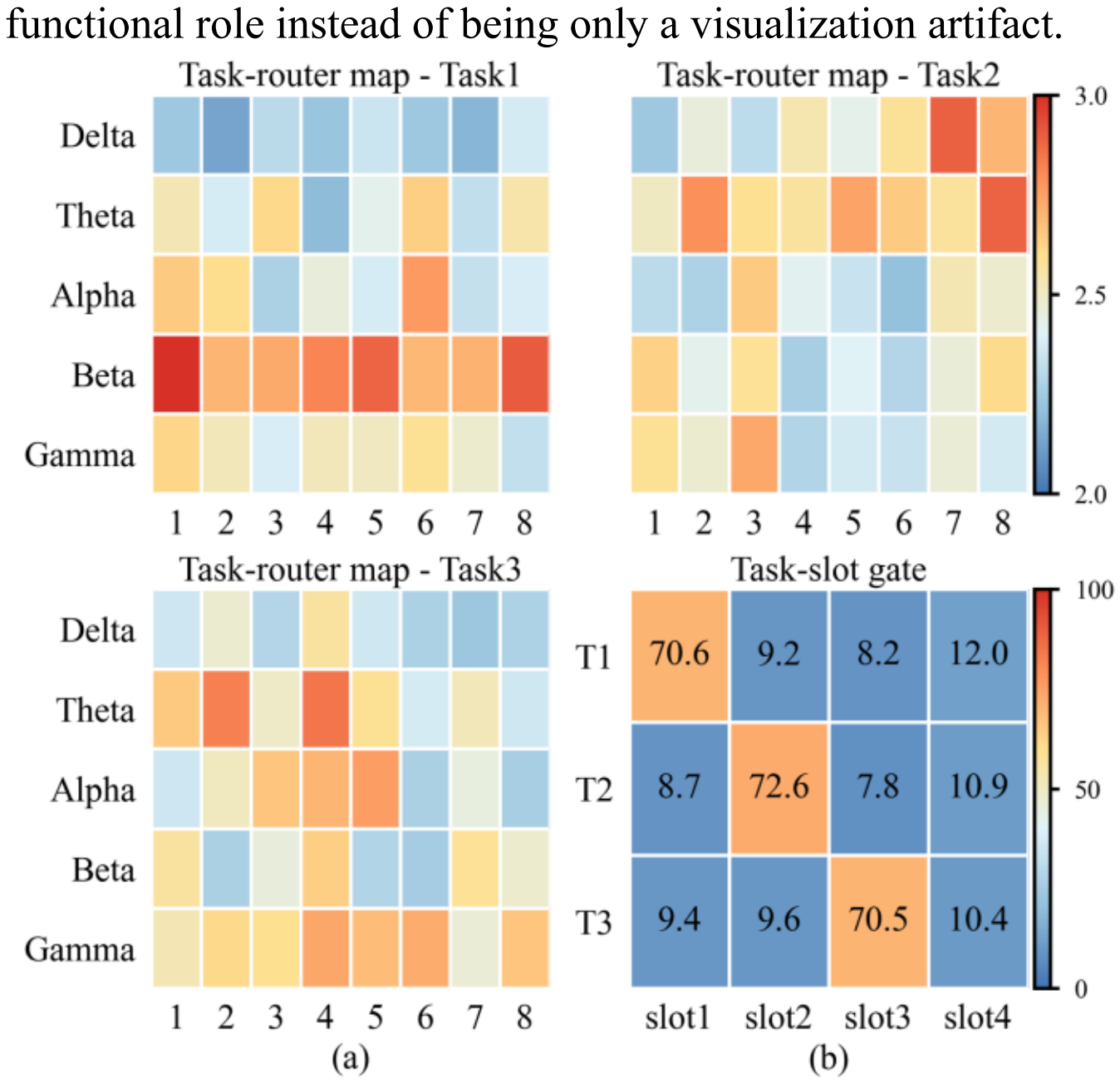

**Fig. 3.** Decomposition analysis of DS-MTNet, averaged across the 20 LOSO folds. (a) Task-router weights (%) over the learned 5 × 8 band-source candidates, one map per task. Rows are the delta, theta, alpha, beta, and gamma bands, and columns 1–8 the source-candidate indices. The color scale is centered at the 2.50% equal-use reference. (b) Task-slot gate weights (%). Each row sums to 100%. T1–T3 are lane-departure detection, steering-response stage decoding, and alertness state decoding.

Overall, the analysis indicates that the performance of DS-MTNet was not explained by one shared backbone alone. The model used a two-level decomposition: the router selected task-dependent mixtures from the band-source dictionary, and the information head assigned each task predominantly to a different reusable slot. The source-router maps gave moderate and descriptive evidence of task-dependent source-frequency routing, while the task-slot gates provided the main and statistically supported evidence for task-specific slot allocation.

### *E. Model-Guided EEG Interpretation*

**Fig. 3** showed that DS-MTNet used task-dependent source routing and task-specific information-slot allocation. A separate question is whether these differently routed pathways were associated with different EEG spectral profiles. To test this, we computed the spectral response of each task-routed source pathway as a source-power contrast between the two conditions of that task in each frequency band, and asked whether this band profile depended on the task. The analysis concerns interpretability and is independent of the decoding contribution of the decomposition, which is established by the ablation (Table III) and the task-slot gates (**Fig. 3b**). **Fig. 4** shows these contrasts over the five source bands for the three tasks, computed the same way for all tasks so that the profiles are directly comparable.

The routed spectral profile depended on the task. A Task × Band analysis showed a significant interaction (permutation $p = 0.005$; Greenhouse-Geisser-corrected $F(8, 152) = 3.24$, $\varepsilon = 0.38$, $p = 0.027$; uncorrected $p = 0.002$), and because this interaction isolates differences in profile shape, the bands that responded varied with the task. The three pairwise task profiles were also distinct (BH-corrected $q < 0.01$ for all pairs), with the largest separation between Task 1 and Task 3. The three task-routed pathways therefore did not reduce to one shared spectral pattern, suggesting that the learned task-routed pathways were associated with different model-guided spectral profiles.

The profiles also differed in shape in a task-appropriate way. For Task 1 (lane-departure epochs minus steady-driving epochs), the routed response was frequency-selective: the alpha and beta contrasts were significantly negative (−1.81 and −1.45 dB; BH-corrected one-sample t-test against zero, $q < 0.001$), whereas the delta contrast was positive and non-significant. Because delta moved opposite to alpha and beta, the Task 1 pathway did not reflect a broadband power shift. Its alpha and beta suppression may be broadly consistent with increased attentional engagement and sensorimotor activation during deviation-related epochs [42].

The other two pathways were directionally consistent but did not reach significance against zero. For Task 2 (steering-response epochs minus preparation epochs), the source-band profile was close to flat, with small reductions that were not significant (the largest was a gamma reduction of −0.44 dB), and this near-flat shape was itself distinct from the other two tasks. For Task 3 (high-RT state minus low-RT state for alertness state decoding), the routed response was negative across all bands, with the largest mean reductions in beta (−1.67 dB) and gamma (−1.90 dB) and with high subject-level variability, and none of these contrasts was significant. Only the Task 1 modulation reached significance against zero, but the between-task profile differences did not depend on each individual profile differing from zero.

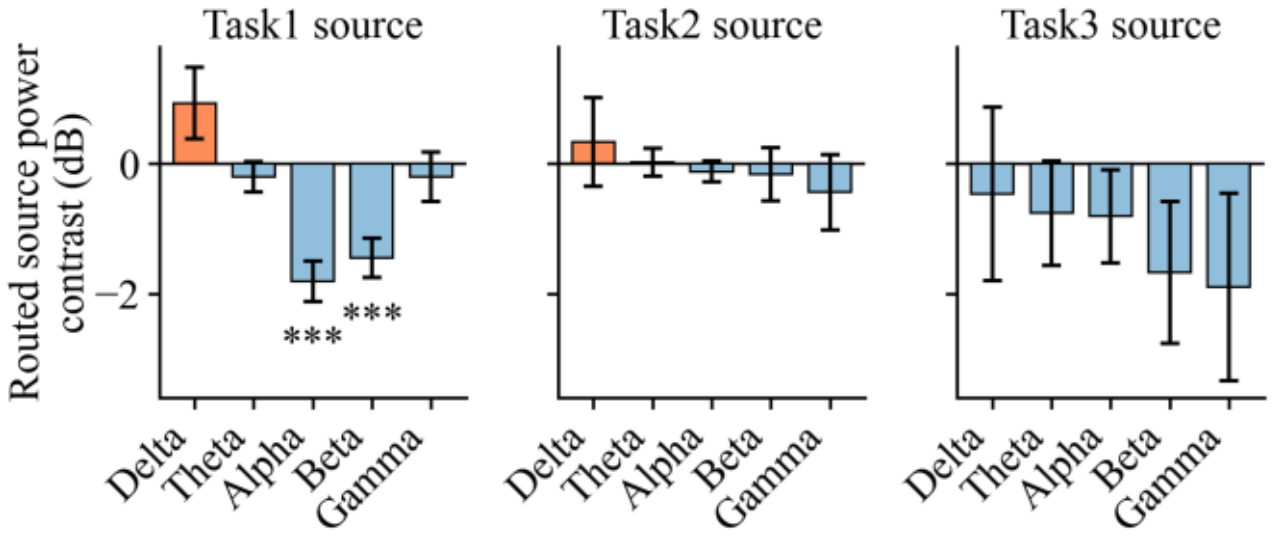

**Fig. 4.** Task-routed source-power contrasts by frequency band. Each panel shows the routed source-power contrast (dB) over the five source bands for one task (Task 1, lane-departure epochs minus steady-driving epochs; Task 2, steering-response epochs minus preparation epochs; Task 3, high-RT state minus low-RT state for alertness state decoding). Bars are subject-level means and error bars are the SEM ($n = 20$). Orange and blue denote positive and negative contrasts. A Task × Band analysis showed a significant interaction (permutation $p = 0.005$; Greenhouse-Geisser-corrected $F(8, 152) = 3.24$, $\varepsilon = 0.38$, $p = 0.027$), and all three pairwise task profiles differed (BH-corrected $q < 0.01$), so the band profile depended on the task. Asterisks mark bands whose contrast differed from zero (BH-corrected one-sample t-test across bands; ***$q < 0.001$).

Overall, **Fig. 4** provides model-guided evidence that the task-routed pathways were associated with different spectral profiles. The significant Task × Band interaction and pairwise profile differences indicate that the three routed pathways did not collapse into one common spectral pattern. The lane-departure pathway additionally showed alpha and beta suppression that was broadly consistent with deviation-related attentional and sensorimotor engagement. The absence of significant against-zero effects for Task 2 and Task 3 suggests that these profiles should be interpreted as model-guided descriptive evidence instead of direct physiological markers. Together with **Fig. 3**, these results support a structured shared-plus-specific representation, in which task readouts use different information slots and routed source-power patterns.

## VI. CONCLUSION

This paper addressed HMC-oriented cognitive sensing by developing DS-MTNet, a structured multi-task EEG decoding framework. The key idea was to organize noisy and mixed windowed EEG activity through frequency-constrained candidate source components, reusable information slots, and task-specific gated readouts. This design enabled a shared model to decode multiple HMC-relevant EEG targets. The driving experiments covered representative information related to environmental events, steering-response stage, and reaction-time-defined alertness state. Experiments on the SADT dataset showed that DS-MTNet achieved the highest mean ACC and F1 among the compared traditional EEG pipelines, single-task deep models, and multi-task baselines under leave-one-subject-out evaluation, with the most statistically robust gains observed for steering-response stage decoding. The ablation results further suggested that the performance gain was associated with the combination of source decomposition, reusable information slots, and task-specific gated readouts, beyond simple parameter sharing alone. The interpretability analyses showed task-specific information-slot readouts and task-dependent routed spectral profiles.

These results support the feasibility of using structured multi-task learning to jointly decode multiple HMC-relevant EEG readouts from task-defined EEG windows. The current study focused on offline EEG decoding and model-guided interpretation, instead of online closed-loop control. A natural next step is to evaluate whether these decoded signals can be integrated into real-time collaborative systems for adaptive warning, task-state inference, or personalized assistance. Future work should also test DS-MTNet across additional HMC scenarios and examine whether the learned information slots remain stable across tasks, subjects, and recording conditions.